\documentclass{article}
\font\sqi=cmssq8
\def\DR{\rm I\kern-1.45pt\rm R}
\def\DC{\kern2pt {\hbox{\sqi I}}\kern-4.2pt\rm C}
\def\DH{\rm I\kern-1.5pt\rm H\kern-1.5pt\rm I}

\newcommand{\ben}{\begin{enumerate}}
\newcommand{\een}{\end{enumerate}}
\newcommand{\beq}{\begin{equation}}
\newcommand{\eeq}{\end{equation}}
\newcommand{\bse}{\begin{subequation}}
\newcommand{\ese}{\end{subequation}}
\newcommand{\bea}{\begin{eqnarray}}
\newcommand{\eea}{\end{eqnarray}}
\newcommand{\bc}{\begin{center}}
\newcommand{\ec}{\end{center}}

\begin{document}
\begin{center}
{\Large\bf Relationship between quantum mechanics
with and without monopoles} \\
\vspace{3mm}

{\large
 Levon Mardoyan$^{1}$, Armen Nersessian$^{1,2}$ and Armen Yeranyan$^{1}$} \\
\end{center}
  {\it ${}^{1}$  Yerevan State University, 1 Alex Manoogian St.,
      Yerevan, 375025,  Armenia\\
      ${}^{2}$ Artsakh State University, 3 Mkhitar Gosh St.,  Stepanakert, Nagorny Karabakh\\
      $\qquad$ Yerevan Physics Institute, 2 Alikhanian Brothers St., Yerevan, 375036, Armenia}

     {\it E-mails: mardoyan@ysu.am ,
  arnerses@yerphi.am , ayeran@ysu.am}
\begin{abstract}
We show that the inclusion of the monopole field in the three- and
five-dimensional spherically symmetric quantum mechanical systems,
supplied by the addition of the special centrifugal term, does not
yield any change in the radial wavefunction and in the functional
dependence of the energy spectra on quantum numbers. The only
change in the spectrum is the lift of the range of the total and
azimuth quantum numbers. The changes in the angular part
wavefunction are independent of the specific choice  of the
(central) potential. We also present  the integrable model of the
spherical oscillator which is different from the Higgs oscillator.
\end{abstract}
\setcounter{equation}{0}
\section{Introduction}
During the last decades there was much activity in the study of
the integrable quantum-mechanical systems specified by the
presence of monopole-like  field configurations. It was initiated
by the pioneer works by Zwanziger \cite{Z} and McIntosh and
Cisneros \cite{mic},  where the analog of Coulomb problem with a
Dirac monopole has been suggested, which inherits whole
(nonlinear) symmetry algebra of the Coulomb system (MICZ-Kepler
system ). The similarity between MICZ-Kepler and Coulomb  systems
has quite transparent explanation in terms of four-dimensional
space: these systems could be obtained from the four-dimensional
oscillator by the reduction by $U(1)$ group action \cite{micred}.
In the same way, one can construct the five-dimensional analog of
the MICZ-Kepler problem, reducing the eight-dimensional oscillator
by $SU(2)$ group action \cite{su2}. In this case, instead of
 Dirac monopole the $SU(2)$ Yang monopole appears in the system.
The uniqueness of these reduction procedures  insists
 on their  close relation with the first and second  Hopf maps (the detailed quantum-mechanical description
 of this correspondence could be found in \cite{ta}).
Let us mention, that the existing analogs of these Coulomb-like
systems on the curved spaces are also specified with the closed
similarity of the systems with and without monopoles
\cite{curvemic}. The actual observable difference  between these
systems results in the lift of the range of the total angular
momentum, which in its turn leads to the degeneracy of the ground
state.

In this paper we show, that the closed similarity between
rotationally invariant three- /five- dimensional
quantum-mechanical systems  with and without Dirac/Yang monopole
is the general peculiarity of these systems. We consider the
quantum mechanics with central potential on the  $d=3,5$-
dimensional  spaces equipped with $so(d)$-invariant metrics \beq
ds^2=g(r)d{\bf r}d{\bf r},\qquad r=|{\bf r}|,\quad {\bf
r}=(x_1,\ldots , x_d)\;. \label{metric}\eeq We shall show, that
incorporation of the monopole supplying with the addition to the
potential of the specific ``centrifugal term", \beq U(r)\to
U(r)+\frac{{\tilde s}^2}{2g(r)r^2}, \label{replacement}\eeq yields
the minor changes of the properties of the system (here ${\tilde
s}^2=\hbar^2 s^2$ with $s$ is the monopole number for the $d=3$
case, and ${\tilde s}^2=\hbar^2 s(s+1)$ with  $s$ is the isospin
of the system for the $d=5$ case). Namely, upon this modification,
the radial wavefunction of the system, as well as the functional
dependence of the energy spectrum on quantum numbers remain
unchanged. The incorporation of the monopole affects the range of
the $SO(d)$ orbital quantum number: it lifts the lowest admissible
value of the orbital quantum number from $0$ to $s$. The angular
part of the wavefunction also changes, but this change is
independent of the specific form of the central potential. In
other words, {\sl for any exactly solvable spherically symmetric
three-/five-dimensional quantum-mechanical system without monopole
we present its explicitly solved generalization with Dirac/Yang
monopole}. Besides the obvious relation with quantum field theory,
these systems could be useful in condensed matter.
``Particle-Dirac monopole" configuration could be used for the
description of the charge in the vicinity of the magnetic pole,
while ``particle-Yang monopole" configuration reduced to low
dimension, could be used for the description of the systems with
spin-orbit interaction. The curved spaces are also related with
condensed matter: they correspond to the systems with effective
non-constant masses. Let us mention, that incorporation of the
monopole, not only provides the system with degenerate ground
state, but also makes possible the dipole transitions preserving
orbital quantum number \cite{tomilchik}. Hence, incorporating the
monopole in the existing quantum dot models, one can provide them
by these specific peculiarities.

The paper is arranged as follows: In {\sl Section 2} we consider
the three-dimensional systems with Dirac monopole, in {\sl Section
3} we consider the five dimensional systems with Yang monopole,
and in {\sl Section 4} we propose the model of the spherical
oscillator which is different from the Higgs oscillator
\cite{higgs}.

\setcounter{equation}{0}
\section{Dirac monopole}
Let us consider  the particle on the space equipped with a $so(3)$-invariant conformal
flat metric (\ref{metric})
moving in the  Dirac monopole magnetic field
 ${\bf B}={\tilde s}{\bf r}/|{\bf r}|^3$ (in the  field)
and in the $so(3)$-invariant potential field
$\tilde{s}^2/2r^2g+U(|{\bf r}|)$. Classical trajectories of this
system are independent of monopole number $\tilde{s}$ \cite{lnp}.
 Let us show, that the same phenomenon takes place in
 the  quantum-mechanical level.
Corresponding quantum-mechanical system is given by the
following Hamiltonian and scalar product
\beq
\hat{\cal H}_s=-\frac{\hbar^2}{2}\triangle_{s}+U(r),\qquad
(\Psi_1,\Psi_2)=\int \psi^*_1 \psi_2\sqrt{\det g}d^3x,
\det g=\det g_{ij}
\label{hamilt}\eeq
The generators of
 $SO(3)$ rotations are defined
by the expressions \beq {\bf J}= {\bf r}\times{\bf p} -{\tilde
s}\frac{{\bf r}}{|{\bf r}|}. \label{jk}\eeq Here \beq
\triangle_s=\frac{1}{2{\det g}^{1/2}}\hat{p}_i g^{ik}{\det
g}^{1/2}\hat{p}_k,\qquad \hat
p_i=-\imath\hbar\frac{\partial}{\partial x_i}-\hbar sA_i, \qquad
s={\tilde s}/\hbar \label{laplas}\eeq where $A_i$ is vector
potential of a Dirac monopole:
$$A_i=\frac{1}{r(r+x_3)}(-x_2,\,x_1,\,0)\;:\; \nabla\times {\bf A}={\bf r}/r^3.$$
Notice, that the momentum operator is not self-conjugated with respect to above scalar product.
However, it could be easily regularized
by the appropriate gauge transformation, preserving the Laplace operator
\beq
{\hat p}_i\to\hat{\bar p}_i=\hat{p}_i-\frac{\imath\hbar}{4{\det g}}\frac{\partial \det g}{\partial
x^i}\;: \quad (\hat{\bar p}_i\Psi_1, \Psi_2)=(\Psi_1,\hat{\bar p}_i \Psi_2),
\eeq
\beq
{\hbar^2}\triangle_{s}=-\frac{1}{\det g^{1/4}}\hat{\bar{p}}_i
g^{ik}\det g^{1/2}\hat{\bar{p}}_k\frac{1}{{\det  g}^{1/4}}=-\frac{1}{{\det g}^{1/2}}\hat{p}_i
g^{ik}{\det  g}^{1/2}\hat{p}_k  .
\eeq
Now, let us specify our formulae for the  metrics
$g_{ij}=g(r)\delta_{ij}$, $\det g_{ij}= g^3$.
Transiting to the spherical coordinates and taking into account the expression 
\beq
\nabla\cdot \mathbf{A}=\frac{1}{\det g^{1/2}}\frac{\partial}{\partial
x^i}(\det g^{1/2}g^{ik}A_k)=0,
\eeq
one could represent the
 Hamiltonian as follows
\begin{equation}
\hat{\cal H}_s=-\frac{\hbar^2}{2} \triangle_r +
\frac{\hat{J}^2}{2g\, r^2}+U\;, \qquad \triangle_r\equiv
\frac{1}{g^{3/2}r^2}\frac{\partial}{\partial
r}\left(g^{1/2}r^2\frac{\partial}{\partial r}\right). \label{h}\end{equation}
Here ${\hat{\bf J}}$ is the quantized angular
momentum of the system, given by (\ref{jk}), so that
\beq
\hat{J}^2=
 -\hbar^2\left[\frac{1}{\sin \theta}\frac{\partial}{\partial \theta}
\left(\sin \theta\frac{\partial}{\partial \theta}\right) +
\frac{1}{\sin^2\theta}\frac{\partial^2}{\partial
\varphi^2}\right]+\frac{2\hbar^2}{1+\cos \theta}\left[s^2-i
s\frac{\partial}{\partial\varphi}\right].\eeq The appropriate
spectral problem reads \beq \hat{\cal
H}_s\Psi_{E,j,m_j}=E\Psi_{E,j,m_j},\quad
\hat{J}^2\Psi_{E,j,m_j}=\hbar^2 j(j+1)\Psi_{E,j,m_j},\quad
\hat{J}_z\Psi_{E,j,m_j}=\hbar m_j\Psi_{E,j,m_j} \label{sspect}\eeq
The variables could be separated by the following choice of the
wavefunction \beq \Psi_{n_r,j,
m_j}(r,\theta,\phi)=\psi_{n_r,j,m_j}(r){\rm
e}^{im_j\phi}d^j_{m_js}(\theta)\;, \eeq where $d^j_{m_js}$ is
Wigner $D$-function. This substitution resolves the last two
equations in (\ref{sspect}), with quantization condition  \beq
j=|s|,|s|+1, \ldots,\qquad m_j=-j,-j+1,\ldots, j-1,j\qquad s=0,\pm
1/2, \pm 1 \ldots \label{jm}\eeq Then  we get the following radial
Schroedinger equation \beq \triangle_r\psi_{n_r,j,m_j}
+\frac{2}{\hbar^2}\left[ E_{n_r,j}-U(r)-\frac{j(j+1)}{g(r)r^2}
\right]\psi_{n_r,j,m_j}=0. \label{radial}\eeq The radial quantum
number $n_r$ is defined from the above radial Schroedinger
equation, and boundary conditions, and depends on the specific
choice of potential $U(r)$. It is seen, that the  radial
Schroedinger equation is independent of the monopole number $s$.
Hence, the spectrum of the Hamiltonian (\ref{hamilt}) remains
unchanged after exclusion of the Dirac monopole.
 The only impact of the
 presence of Dirac monopole in the spectrum  is in the range of definition of the orbital and
 azimuth quantum numbers $j,m_j$ given by (\ref{jm}).
 The impact of the Dirac monopole in the wavefunction concerns   spherical part, while
 radial wavefunction remains unchanged.

\setcounter{equation}{0}
\section{Yang monopole}
Now let us consider  the five-dimensional $SO(5)$-invariant quantum mechanics with the
$SU(2)$ Yang monopole.
The Hamiltonian of this system  is  again given by
the expression (\ref{hamilt}), where the momenta operators
$\hat{p}_i$ in (\ref{laplas})
are replaced by the following ones
\begin{equation}
\hat{P}_i=\hat{\bar{p}}_i -\hbar A^a_i \hat{T}_a,\qquad a=1,2,3, \quad
\left[\hat{T}_a,\,\hat{T}_b\right]=\imath\varepsilon_{abc}\hat{T}_c\;.
\end{equation}
The  five-dimensional vector potential $\mathbf{A}^a$ is defined
by the expressions
\begin{eqnarray}
\mathbf{A}^1 =\frac{1}{r(r+x_5)}( x_4,x_3,-x_2,-x_1,0), \quad
\mathbf{A}^2 =\frac{1}{r(r+x_5)}(-x_3, x_4,-x_1,-x_2, 0),\quad
\mathbf{A}^3 =\frac{1}{r(r+x_5)}(x_2,-x_1, x_4,- x_3, 0)
\end{eqnarray}
and obey the equations
\begin{equation}
\mathbf{A}^a\cdot\mathbf{A}^b= \frac{1}{r^2}\frac{r- x_5}{r+
x_5}\delta^{ab}, \quad \mathbf{A}^a\cdot\mathbf{r}=0,\quad \nabla
\cdot \mathbf{A}^a =0
\end{equation}
This potential describes the Yang monopole with topological charge
$+1$. We shall restrict ourselves by the detailed consideration of
this case, since the transition to the system with the Yang
anti-monopole (i.e. the Yang monopole with topological charge
$-1$) is straightforward.

Let us pass to the
 five-dimensional Euler coordinates
\beq
x_5=r\cos\theta, \quad
x_1+\imath x_2=r\sin\theta
\cos\frac{\beta}{2}{\rm e}^{\imath\frac{\alpha+\gamma}{2}},
\quad x_3 +\imath x_4=r\sin\theta
\sin\frac{\beta}{2}{\rm e}^{\imath\frac{\alpha-\gamma}{2}}.
\eeq
In these coordinates  the Hamiltonian looks as follows
\begin{eqnarray}\label{ham5}
\hat{H}=-\frac{1}{2}\hbar^2\frac{1}{g^{5/2}r^4}\frac{\partial}{\partial
r}\left(g^{3/2}r^4\frac{\partial}{\partial r}\right)-\frac{\hbar^2{\hat\Lambda}^2}{g
r^2 } +U,\qquad \hat{\Lambda}^2\equiv -\frac{1}{ \sin^3
\theta}\frac{\partial}{\partial \theta} \left(\sin^3
\theta\frac{\partial}{\partial \theta} \right) +\frac{\hat{L}^2}{
\sin^2(\theta/2)}+\frac{\hat{J}^2}{
\cos^2(\theta/2)}\;.
\end{eqnarray}
Here  $\hat{L}_a$ and $\hat{J}_a$ are the generators of
$so(4)=so(3)\times so(3)$ algebra, \beq
\hat{J}_a=\hat{L}_a+\hat{T}_a\;:\quad
\left[\hat{L}_a,\,\hat{T}_b\right]=0,\quad
\left[\hat{L}_a,\,\hat{L}_b\right]=\imath\varepsilon_{abc}\hat{L}_c,\quad
\left[\hat{J}_a,\,\hat{J}_b\right]=\imath\varepsilon_{abc}\hat{J}_c
\eeq and ${\hat \Lambda}^2$ defines the total  $SO(5)$ momentum of
the system. Explicitly the generators ${\hat L}_a$ look as follows
\begin{eqnarray}
\hat{L}_1=\imath \left(\cos \alpha \cot
\beta\frac{\partial}{\partial \alpha}+ \sin
\alpha\frac{\partial}{\partial \beta}-\frac{\cos \alpha}{\sin
\beta}\frac{\partial}{\partial \gamma}\right),
\quad\hat{L}_2=\imath \left(\sin \alpha \cot
\beta\frac{\partial}{\partial \alpha}- \cos
\alpha\frac{\partial}{\partial \beta}-\frac{\sin \alpha}{\sin
\beta}\frac{\partial}{\partial
\gamma}\right),\quad\hat{L}_3=-\imath\frac{\partial}{\partial
\gamma}\;. \label{angm}\end{eqnarray} The dimensions of the
quantum mechanics with and without Yang monopole are different
ones,
 because of non-Abelian nature of the Yang monopole.
 Namely, in the absence of the Yang  monopole the system is five-dimensional one, and its wavefunction
 depends on ${\bf r}$ coordinates only. When we incorporate
 in the system the  Yang monopole, we should take into account its internal space,
the  two-dimensional sphere $S^2$ (the maximal orbit of $SU(2)$ group).
 However, it is more convenient to consider the three-dimensional sphere $S^3$ instead of
 $S^2$.
 For this purpose we define
 the generators ${\hat T}_a$ in terms of the angles $\alpha_T,\beta_T, \gamma_T$,
 parameterizing three-dimensional sphere $S^3$. The explicit expressions are given by
 (\ref{angm}), where ${\hat L}_a$
 are replaced by ${\hat T}_a$, and $\alpha, \beta,\gamma$ by the $\alpha_T,\beta_T, \gamma_T$.
 In that case one can consider the wavefuncion depending on coordinates
 $(r,\theta, \alpha, \beta, \gamma,\alpha_T,\beta_T, \gamma_T)$, with the additional constraint
 imposed by
\beq {\widehat T}^2\Psi(r,\theta, \alpha, \beta,
\gamma,\alpha_T,\beta_T, \gamma_T )=s(s+1)\Psi(r,\theta, \alpha,
\beta, \gamma,\alpha_T,\beta_T, \gamma_T ), \eeq
where  $s$ is
positive integer defining the isospin of the particle. So, after
inclusion of the  Yang monopole the initial five-dimensional
system becomes six-dimensional one. Now, let us introduce the
separation ansatz \beq \Psi(r,\theta,\alpha,\beta,\gamma,
\alpha_T,\beta_T,\gamma_T) = R(r)Z(\theta
)\Phi(\alpha,\beta,\gamma,\alpha_T,\alpha_T, \gamma_T).
\label{sep}\eeq which resolves
 the following spectral problem
\bea
&{\cal H}\Psi={\cal E}\Psi, \quad {\hat\Lambda}^2\Psi=\Lambda (\Lambda+3)\Psi\;,&\label{5s1}\\
& {\hat L}^2\Psi= L(L+1)\Psi,\quad {\hat
T}^2\Psi={s(s+1)}\Psi,\quad {\hat J}^2\Psi={J(J+1)}\Psi,\quad
{\hat J}_3\Psi=M\Psi.&\label{5s2} \eea The functions  $\Phi$  are
the eigenfunctions of ${\hat L}^2$, ${\hat T}^2$,${\hat J}^2$,
${\hat J}_3$ with the eigenvalues $L(L+1)$, $s(s+1)$, $J(J+1)$ and
$M$ respectively. Hence, $\Phi$ could be represented in the form
\beq \Phi =
\sum_{M=m+t}\left(JM|L,m';s,t'\right)D_{mm'}^L(\alpha,\beta,\gamma)
D_{tt'}^s(\alpha_T,\beta_T,\gamma_T) \eeq where
$\left(JM|L,m';s,t'\right)$ are the Clebsh-Gordan coefficients and
$D_{mm'}^L$ and $D_{tt'}^s$ are the Wigner functions. These
quantum numbers have the following range of definition \beq
L=0,1,\ldots \;;\qquad J=|L-s|,|L-s|+1,\ldots,|L+s|-1,|L+s|\;;
\qquad M=-J,-J+1,\ldots, J-1, J. \eeq The function $Z(\theta )$ is
the eigenfunction of the $SO(5)$ momentum operator: \beq
{\hat\Lambda}^2Z(\theta)=\Lambda(\Lambda+3)Z(\theta)\;:\quad
Z_{\Lambda L J}=C_{\Lambda L
J}(1-\cos\theta)^{L}(1+\cos\theta)^{J}P_{\Lambda-L-J}^{(2L+1,2J+1)}(\cos\theta),
\eeq where $C_{\Lambda L J}$ is normalization constant, \beq
C_{\Lambda L
J}=\sqrt{\frac{(2\Lambda+3)(\Lambda-J-L)!\Gamma(\Lambda+J+L+3)}{2^{2J+2L+3}
\Gamma(\Lambda+J-L+2)\Gamma(\Lambda-J+L+2)}}, \eeq and
$P_{\Lambda-L-J}^{(2L+1,2J+1)}$ are Jacobi polynomials. Hence, we
get the following quantization condition \beq \Lambda=
n_\theta+L+J,\quad n_\theta=0,1,\ldots\quad . \eeq Substituting
(\ref{sep}) into (\ref{ham5}) and taking into account appropriate
eigenvalues we get  the equation for the radial  wavefunction
 \beq \frac{1}{g^{5/2}r^4}\frac{d}{d
r}\left(g^{3/2}r^4\frac{d R}{d
r}\right)-\frac{\Lambda(\Lambda+3)}{g r^2}R+\frac{2}{\hbar^2}
(E-U)R=0.\eeq It is seen that the radial Schroedinger equation is
independent of the isospin $s$, so that the impact of the Yang
monopole in the spectrum results in the change of the range of
quantum number $\Lambda$ (and of $J$ and $M$ as well).
 The impact of the Yang monopole in the wavefunction concerns   spherical part and is independent of the
 specific form of the potential, while
 radial wavefunction remains unchanged.

Consideration of the system with Yang anti-monopole (Yang monopole
with $-1$ topological charge) is completely similar to the above
one. The respective formulae can be obtained by the given ones by
the replacement \beq J\to L,\quad L\to J. \eeq Hence, we get that
the consequences of the inclusion of the Yang (anti-)monopole   in
the five-dimensional $SO(5)$ symmetric system supplemented with
the appropriate change of potential
 given by (\ref{replacement}), is completely similar to the inclusion of Dirac monopole in the
 three-dimensional $SO(3)$-symmetric system.

\setcounter{equation}{0}
\section{Spherical oscillators}
In previous Sections we proposed the explicit expressions for the
wavefunctions and spectra of the generalizations of three-/five-
dimensional spherically symmetric exactly solvable quantum
mechanical systems (without external gauge fields). Clearly, our
construction includes, as particular cases, the  MICZ-Kepler
systems on the  three- and five- dimensional Euclidean spaces,
spheres  and hyperboloids (see second and third references in
\cite{curvemic}, as well as Ref. \cite{meng}, where the
MICZ-Kepler system on the any-dimensional spheres and hyperboloid
has been constructed). For example, in conformal flat coordinates
the metric of sphere and two-sheet hyperboloid look as follows
\beq g{d{\bf r}^2}=\frac{4 r^2_0d{\bf r}^2}{(1+\varepsilon
r^2)^2},\quad \varepsilon=\pm 1 \eeq where $\varepsilon=1$
corresponds to the sphere, and $\varepsilon=-1$ corresponds to the
(two-sheet) hyperboloid, and $r_0$ is the radius of the sphere
(hyperboloid). In these coordinates the potential of the Higgs
oscillator reads \beq
U_{Higgs}=\frac{\omega^2r^2_0}{2}\frac{4r^2}{(1-\varepsilon
r^2)^2}. \eeq The energy spectrum of the Higgs oscillator spectrum
of this system is given by the expression \cite{pogos} \beq
E_{n_r, \Lambda} = \frac{1}{2r^2_0}\left[\left(2\sqrt{\omega^2
r^4_0+\frac14}-1\right)\left(2n_r+\Lambda + \frac{d}{2}\right)
+\varepsilon\left(2n_r+\Lambda
+\frac{\varepsilon+1}{2}\right)\left(2n_r+\Lambda
+d+\frac{\varepsilon -1}{2}\right) \right]. \eeq Here $\Lambda$
denotes the orbital quantum number, reducing to $\Lambda=j$ for
$d=3$. The radial quantum number $n_r$ takes the values
$n_r=0,1,\ldots $ for $\varepsilon=1$, and $n_r=0,1,\ldots ,
[\sqrt{\omega^2 r^4_0+\frac14}-d/2-\Lambda/2]/2$ for
$\varepsilon=-1$. Incorporation of the monopole yields the change
of the range of $\Lambda$  from $\Lambda=0,1,\ldots$ to
$\Lambda=s,s+1, \ldots$.
\\

Let us conclude this Section proposing the alternative model of
the spherical oscillator. It was initially suggested as a model of
oscillator  on the complex projective space $\DC P^N$ and
quaternionic projective space $\DH P^N$ \cite{cpn} respecting the
inclusion of constant magnetic field (on $\DC P^N$) and BPST
instanton field (on $\DH P^N$)\cite{hpn}. For the $N=1$, i.e. $\DC
P^1=S^2$ and $\DH P^N= S^4$ the corresponding potentials are
defined by the expression \beq V({\bf
r})={2\omega^2r^2_0r^2}=2\omega^2r^2_0\frac{1-\cos\chi}{1+\cos\chi},
\label{np}\eeq where $\chi$ is the spherical coordinate of the
respective sphere, $x_d=r_0\cos\chi$.

Surprisingly, this potential defines the integrable generalization
of the oscillator in the arbitrary-dimensional spheres too.
Indeed, with the formulae for  the Schroedinger equation of the
spherically-symmetric system on the $d$-dimensional sphere at hand
\cite{pogos}, one could immediately get the energy spectrum and
the wavefunctions for the system with potential. Namely, the
spectrum of the system is defined by the expression \beq
E_{n_r,\Lambda}=\frac{1}{2r^2_0}\left[(2n_r+\Lambda+1)(2n_r+\Lambda+d)+
(2\nu-1)(2n_r+\Lambda+\frac{d}{2}) \right], \qquad
\nu^2=(\Lambda+\frac{d-2}{2})^2+16\omega^2r^4_0 \eeq where
$n_r=0,1,\ldots $ is radial quantum number, and
$\Lambda=0,1,\ldots$
 is the $SO(d+1)$ orbital quantum number.
 The radial wavefunction looks as follows
\beq
R^D_{n_r \Lambda\nu}(\chi)=
C_{n_r\Lambda\nu}
(\sin\chi )^\Lambda
(\cos\chi)^{\nu -d/2+1}P^{(\Lambda+d/2 -1,\nu)}_{n_r}(\cos\chi)
\eeq
where $P^{(a,b)}_n$ is Jacobi polinomial
and $C_{n_r\Lambda\nu}$ is normalization constant
\beq
C_{n_r\Lambda\nu}=\frac{1}{2^{(d-1)/2}}
\sqrt{\frac{(2n_r+\Lambda+\nu+d/2){n_r}!\Gamma(n_r+\Lambda+\nu+d/2)}{{r^d_0}\Gamma(n_r+\Lambda+d/2)
\Gamma (n_r+\nu+1)}}\;.
\eeq
One can expect, that this oscillator model will respect the inclusion of the $SO(d)$ instanton field
not only in $d=2,4$, but in any  $d$ too. It is  also clear, that the same potential
will define the integrable system on the $d$-dimensional two-sheet hyperboloid.

 \section*{Conclusion}
 We have shown, that the incorporation of the Dirac/Yang monopole in the spherically symmetric
 exactly solvable three-/five-dimensional quantum mechanical system (without external gauge fields),
 supplemented by the change of the potential given by
 Eq. (\ref{replacement}), yields the exactly solvable generalization of the initial system.
Moreover, we have shown, that the radial  wavefunction remains
unchanged upon given modification, as well as the functional
dependence of the energy spectrum on radial orbital and azimuth
quantum numbers. Hence, having at hand the exact solution of the
quantum-mechanical system without monopole, we can immediately
present the exact solutions of the respective system with
monopole. We suppose, that the similar correspondence takes place
not only in $d=3,5$ dimensional spherically symmetric systems, but
for the arbitrary $d$-dimensional $SO(d)$-symmetric system, upon
inclusion of  $SO(d-1)$ monopole field. Also, we present some
exactly solvable model of the oscillator on the $d$-dimensional
sphere: besides the $SO(d-1)$ monopole field, this model would,
presumably, respect the inclusion of $SO(d)$ instanton field too.
Such a monopole and instanton  solutions are given in
\cite{tigran}.

The similar analysis of the systems with axial symmetry is more
complicated but more important (the simplest system of this sort,
specified with the presence of Dirac monopole was proposed  in
\cite{gmard}). Indeed, presented way of the incorporation of the
monopole in the spherically symmetric systems does not yield
essential change in the system's properties. However, their
interaction with the external fields (without spherical symmetry)
could yield qualitatively different effects,
 as it was observed in the study of the
Stark effect in the MICZ-Kepler system \cite{stark}.\\

{\Large Acknowledgments.}  The work has been supported in part by  the grants
NFSAT-CRDF  ARP1-3228-YE-04 and INTAS-05-7928 .


\begin{thebibliography}{100}

\bibitem{Z} D.~Zwanziger,
  Phys.\ Rev.\  {\bf 176} (1968) 1480.
\bibitem{mic}
H. ~McIntosh, A. ~Cisneros. 
 J. Math. Phys., {\bf 11} (1970) 896.

\bibitem{micred}
T. Iwai, Y. Uwano, 
J. Math. Phys. {\bf 27}(1986), 1523.

I.M. Mladenov and V.V. Tsanov, 
J. Phys., {\bf A20} (1987), 5865;
 J. Phys., {\bf
A32} (1999), 3779.

%

\bibitem{su2}T.~Iwai, 
 J.~Geom.~Phys.~{\bf 7}~(1990),~507;

L.~G.~Mardoyan, A.~N.~Sissakian, V.~M.~Ter-Antonyan, 
 Phys.~Atom.~Nucl.~{\bf 61}~(1998),~1746

 \bibitem{ta}  V.~Ter-Antonian,
[arXiv:quant-ph/0003106]
 {\sl Lectures given at BLTP International School on Symmetries and Integrable Sysems,
  Dubna, Russia, 8-11 June 1999}

 L.~G.~Mardoyan, G.~S.~Pogosyan, A.~N.~Sissakian and V.~M.~ Ter-Antonyan,
 {\sl Quantum systems with hidden symmetry}, MAIK Publ., Moscow, 2006



\bibitem{curvemic} G.~W.~Gibbons and N.~S.~Manton,
  Nucl.\ Phys.,\  B {\bf 274}, 183 (1986);

V.~V.~Gritsev, Yu.~A.~Kurochkin, V.~S.~Otchik,
 J.~Phys.~A: Math.~Gen.{\bf 33}(2000), 4903

  A.~Nersessian and G.~Pogosyan,
  Phys.\ Rev.\ A {\bf 63} (2001) 020103


  S.~Bellucci, A.~Nersessian and A.~Yeranyan,
  Phys.\ Rev.\ D {\bf 70} (2004) 045006


  G.~W.~Gibbons and C.~M.~Warnick,
  arXiv:hep-th/0609051.

%
%
%
%
%
%
%
%
%
%
%
\bibitem{tomilchik}
 E.~A.~Tolkachev, L.~M.~Tomilchik and Y.~M.~Shnir,
  Yad.\ Fiz.\  {\bf 38}(1983), 541;
  J.\ Phys.\ G {\bf 14} (1988) 1.

\bibitem{lnp} A.~Nersessian,
 Lect. Notes Phys. {\bf 968} (2006), 139 [arXiv:hep-th/0506170].

\bibitem{meng}
G.~w.~V.~Meng, 
[arXiv: math-ph/0507028]
\bibitem{higgs} P.~W.~Higgs,
  J.\ Phys.\ A {\bf 12} (1979) 309.;

 H.~I.~Leemon,
  J.\ Phys.\ A {\bf 12} (1979) 489.
\bibitem{pogos}
  E.~G.~Kalnins, W.~.~J.~Miller and G.~S.~Pogosyan,
  Phys.\ Atom.\ Nucl.\  {\bf 65}, 1086 (2002)

\bibitem{cpn}
  S.~Bellucci and A.~Nersessian, 
  Phys.\ Rev.\ D {\bf 67} (2003) 065013

  S.~Bellucci, A.~Nersessian, A.~Yeranyan,
  Phys.\ Rev.\ D {\bf 70} (2004) 085013

\bibitem{hpn}
  L.~Mardoyan and A.~Nersessian,
  Phys.\ Rev.\ B {\bf 72} (2005), 233303

  S.~Bellucci, L.~Mardoyan and A.~Nersessian,
  Phys.\ Lett.\ B {\bf 636}, 137 (2006) 
\bibitem{tigran}
  D.~H.~Tchrakian,
  J.\ Math.\ Phys.\  {\bf 21} (1980) 166.

  \bibitem{gmard}
  L.~Mardoyan,
  J.\ Math.\ Phys.\  {\bf 44} (2003), 4981;
  Phys.\ Atom.\ Nucl.\  {\bf 68}, 1746 (2005)
  [arXiv:quant-ph/0310143].

\bibitem{stark}
  L.~Mardoyan, A.~Nersessian and M.~Petrosyan,
  Theor.\ Math.\ Phys.\  {\bf 140}, 958 (2004)
\end{thebibliography}
\end{document}